\documentclass[12pt,a4paper]{article}
\usepackage{latexsym}
\font\mybb=msbm10 at 12pt
\def\bb#1{\hbox{\mybb#1}}

\def\unit{\hbox to 3.3pt{\hskip1.3pt \vrule height 7pt width .4pt \hskip.7pt
\vrule height 7.85pt width .4pt \kern-2.4pt
\hrulefill \kern-3pt
\raise 4pt\hbox{\char'40}}}
\def\II{{\unit}}

\def\bb#1{\hbox{\mybb#1}}

\def\bE{\bb {E}}

\def\dalemb#1#2{{\vbox{\hrule height .#2pt
        \hbox{\vrule width.#2pt height#1pt \kern#1pt
                \vrule width.#2pt}
        \hrule height.#2pt}}}

\let\u=\upsilon

 \def\bd{\begin{document}} \def\ed{\end{document}}
\def\ds{\documentstyle} \let\fr=\frac \let\bl=\bigl \let\br=\bigr
\let\Br=\Bigr \let\Bl=\Bigl
\let\bm=\bibitem
\let\na=\nabla
\let\pa=\partial \let\ov=\overline
\newcommand{\be}{\begin{equation}}
\newcommand{\ee}{\end{equation}}
\def\ba{\begin{array}}
\def\ea{\end{array}}
\def\ft#1#2{{\textstyle{{\scriptstyle #1}\over {\scriptstyle #2}}}}
\def\fft#1#2{{#1 \over #2}}
\def\del{\partial}
\def\ui{{\underline i}}
\def\uj{{\underline j}}
\def\uk{{\underline k}}
\def\u0{{\underline 0}}
\def\ur{{\underline r}}
\def\ul{{\underline \ell}}
\def\url{{\underline {r+\ell}}}
\def\sst#1{{\scriptscriptstyle #1}}
\def\oneone{\rlap 1\mkern4mu{\rm l}}
\newcommand{\ho}[1]{$\, ^{#1}$}
\newcommand{\hoch}[1]{$\, ^{#1}$}
\newcommand{\bea}{\begin{eqnarray}}
\newcommand{\eea}{\end{eqnarray}}
\newcommand{\ra}{\rightarrow}
\newcommand{\lra}{\longrightarrow}
\newcommand{\Lra}{\Leftrightarrow}
\newcommand{\ap}{\alpha^\prime}
\newcommand{\bp}{\tilde \beta^\prime}
\newcommand{\tr}{{\rm tr} }
\newcommand{\Tr}{{\rm Tr} }
\newcommand{\NP}{Nucl. Phys. }

\begin{document}

\begin{flushright}
\footnotesize
Groningen/UG-4/97\\
Stanford/97-17\\
CERN--TH/97--84\\
DAMTP/R/97/17\\
\hfill{\bf hep-th/9705040}\\
May 8th 1997
\normalsize
\end{flushright}

\vspace{10pt}

\begin{center}

{\Large {\bf  $\kappa$-Symmetry, Supersymmetry and Intersecting Branes}}

\vspace{20pt}

{\bf E.~Bergshoeff}

{\it Institute for Theoretical Physics, Nijenborgh 4, 9747 AG Groningen,\\
The Netherlands}

\vspace{10pt}

{\bf R.~Kallosh}

{\it Department of Physics, Stanford University, Stanford, California 94305, 
USA}

\vspace{10pt}

{\bf T.~Ort\'{\i}n}\hoch{\sst\star}

{\it C.E.R.N.~Theory Division, CH--1211, Gen\`eve 23, Switzerland}

\vspace{10pt}

{\bf G.~Papadopoulos}

{\it DAMTP, University of Cambridge, Silver Street, Cambridge, CB3
9EW, UK}

\vspace{15pt}

\underline{ABSTRACT}

\end{center}

\begin{quote}
\small

We present a new form of $\kappa$-symmetry transformations for
D-branes in which the dependence on the Born-Infeld field strength is
expressed as a relative rotation on the left- and right-moving
fields with opposite parameters.  Then, we apply this result to
investigate the supersymmetry preserved by certain intersecting brane
configurations at arbitrary angles and with non-vanishing constant
Born-Infeld fields. We also comment on the covariant quantization of
the D-brane actions.

\normalsize

\end{quote}

\vspace{.5cm}

\begin{flushleft}
\footnotesize
CERN--TH/97--84
\normalsize
\end{flushleft}

{\vfill\leftline{}\vfill
\footnoterule
{\footnotesize
        \hoch{\sst\star} Address after October 1997:  
       {\it IMAFF, CSIC, Calle de Serrano 
            121, E-28006-Madrid, Spain}
\vskip  -12pt} \vskip 10pt
\newpage

\setcounter{page}{1}


\section{Introduction}

It is well-known that the covariant formulation of superstrings
\cite{grsh} and supermembranes \cite{begpkt} is based upon a special
fermionic gauge symmetry on the worldvolume which is called
$\kappa$-symmetry.  Upon gauge-fixing this $\kappa$-symmetry, the
global target space supersymmetry combines with a special
field-dependent $\kappa$-transformation into a global worldvolume
supersymmetry. This worldvolume supersymmetry guarantees the equality
of bosonic and fermionic degrees of freedom on the worldvolume. This
close relationship between $\kappa$-symmetry and supersymmetry can be
applied to determine the fraction of spacetime supersymmetry preserved
by certain single bosonic string and membrane configurations (see,
e.g.  \cite{BDPS}).  A recent development has been the construction of
$\kappa$-symmetric non-linear effective actions and/or equations of
motion for D-branes \cite{Ce1,Ag1,Be1,Ho1} and the M-5-brane
\cite{Ho2,Ba1,Ag2} complementing the $\kappa$-symmetric superstring
and M-2-brane \cite{grsh,begpkt}.  In these new cases, there is again
a close relation between $\kappa$-symmetry and supersymmetry which
leads to an equality of bosonic and fermionic degrees of freedom. We
will apply this relation to investigate the supersymmetry preserved by
certain single bosonic D-brane and M-brane configurations.

Apart from the single-brane configurations, in many applications of
superstring dualities a central role is played by intersecting-brane
configurations that preserve an, in general smaller, fraction of the
vacuum spacetime supersymmetry. The allowed intersections depend on
the worldvolume field content of the branes involved in the
intersection \cite{gppkt}.  The effective action of an intersecting
configuration is expected to be a non-Abelian generalisation of the
single brane actions.  In the linearised limit, this action becomes
that of a coupled system with (non-Abelian) vector, tensor and matter
multiplets.  For example, if the branes involved in the intersection
are D-branes, the effective theory is a Yang-Mills theory coupled to
matter.  In the ``Abelian'' limit the effective action of an
intersecting brane configuration reduces to a non-linear action
similar to that of a single brane.

All known $\kappa$-symmetry transformations of brane actions take the
form

\begin{equation}
\delta\theta= (1+\Gamma) \kappa ,
\label{inone}
\end{equation}

\noindent where $\theta$ is a spacetime spinor depending on the 
worldvolume coordinates $\sigma$, $\kappa(\sigma)$ is the parameter of
the $\kappa$-transformation and $\Gamma$ is a hermitian traceless
product structure i.e.

\begin{equation}
{\rm tr}\ \Gamma=0\, ,
\hspace{1cm}
\Gamma^2=1\, .
\end{equation}

\noindent The expression for $\Gamma$ depends on the embedding map $X$
from the worldvolume of the brane into spacetime, and for D-branes is
non-linear in

\begin{equation}
\label{calF}
{\cal F} =  F - B\, ,
\end{equation}

\noindent where $F$ is the Born-Infeld (BI) 2-form field strength 
and $B$ is the background NS/NS two-form gauge potential.

In this paper we shall show that the non-linear 
dependence of $\Gamma$ on ${\cal F}$
can be expressed as

\begin{equation}
\label{agamma}
\Gamma= e^{-a/2}\ \Gamma_{(0)}^{\prime}\ e^{a/2}\, ,
\end{equation}

\noindent where $a = a({\cal F})$ contains all the dependence 
on the BI field and $\Gamma^{\prime}_{(0)}$ (which depends only on
$X$) is also a hermitian traceless product structure (i.e.~tr
$\Gamma_{(0)}^{\prime} = 0$ and $\left(\Gamma_{(0)}^{\prime}\right)^2
= 1$).  In this new form of $\Gamma$, the proof that $\Gamma$ is a
hermitian traceless product structure is straightforward.  As another
application we shall use (\ref{agamma}) to investigate the
supersymmetry preserved by intersecting brane configurations.

The classical D-brane actions have in addition to worldvolume
$\kappa$-symmetry also a 32-component spacetime supersymmetry.  As we
shall see for single bosonic brane-probe configurations\footnote{In
  this paper we define brane probes as solutions of the worldvolume
  action for fixed target space background.}, the fraction of the
supersymmetry preserved is determined by the number of solutions of
the following equation\footnote{The same condition has been derived in
  the boundary state formalism \cite{BDL,BL}.}:

\begin{equation}
(1-\Gamma) \epsilon=0,
\label{intwo}
\end{equation}

\noindent where $\epsilon$ is the spacetime supersymmetry parameter. 
For brane probes this is the only supersymmetry condition that arises.
However, for supergravity configurations with branes as sources (that
is, for BI D-p-brane actions coupled to the supergravity action), the
above condition must be complemented with the usual Killing spinor
equation of the supergravity theory.  In all cases that we know of the
supergravity Killing spinor equation implies the above
condition\footnote{Apparently supersymmetric solutions always have
  supersymmetric sources. It would be interesting to have a general
  and rigorous proof of this.}.

Several methods can be used to find the fraction of supersymmetry
preserved by intersecting brane configurations.  In this paper we
shall apply (\ref{intwo}) to investigate the supersymmetry preserved
by such configurations.  For this we shall introduce the projection
(\ref{intwo}) for each brane involved in the intersection and then we
shall examine the compatibility of all the projections. One of the
advantages of this method is that all intersections can be treated in
a unified way.  To simplify the computation, we shall first assume
that all the branes involved in the intersection are probes
propagating in the D=10 Minkowski spacetime.  In this case, we shall
find that one can take the BI fields associated with the D-branes and
M-branes to be constant rather than zero. For vanishing BI fields, we
shall reproduce all the known results for the allowed supersymmetric
intersecting brane configurations.

Next we shall briefly
comment on D-branes in their appropriate supergravity background.
The matching of the supergravity solution to the
source necessitates that the BI field of the brane must vanish if the
supergravity solution does not contain a non-vanishing NS/NS two-form 
gauge potential.

There are some limitations to the above method for determining the
fraction of supersymmetry preserved by intersecting brane
configurations. One is that we are considering Abelian BI-type
effective actions despite the fact that the full effective theory is
expected to be non-Abelian. The non-Abelian case corresponds to
configurations of coincident branes. Such configurations will not be
considered in this paper.  We have also ignored parts of the effective
action; for example in intersections involving D-branes, we have not
taken into account the matter multiplets that are associated with open
strings ending at two different D-branes involved in the intersection.
Nevertheless, the results of our paper apply in the ``Abelian'' limit
of the full theory.

The organization of this paper is as follows: In section two, we shall
review the action and $\kappa$-symmetry transformations of D-p-branes.
In section three we derive the new form of the
$\kappa$-transformations given in eqs.~(\ref{inone}) and
(\ref{agamma}).  In section four we discuss the condition
(\ref{intwo}) for supersymmetric configurations. In section five we
shall investigate the conditions for intersecting D-brane probes to
preserve a fraction of spacetime supersymmetry.  In section six, we
extend our results to include M-branes. In section seven we comment on
the supersymmetry preserved by supergravity/brane configurations and
in appendix A, we shall comment on the covariant quantization of
D-brane actions.


\section{D-Branes and $\kappa$-Symmetry}

To make our discussion self-contained we briefly review here the
basics of $\kappa$-symmetry. For a more detailed discussion and our
notation we refer to \cite{Be1}. Let $G, B$ and $\phi$ be the
spacetime metric, the NS/NS 2-form gauge potential and the dilaton,
respectively.  The bosonic D-p-brane is described by a map $X$ from
the worldvolume $\Sigma_{(p+1)}$ into the $d=10$ spacetime ${\cal M}$
and by a 2-form BI field strength $F$ on $\Sigma_{(p+1)}$; $dF=0$ so
$F=dV$ where $V$ is the one-form BI gauge potential. The bosonic part
of the effective action of a D-p-brane is

\begin{equation} 
\label{action}
I_{\rm p} = -\int d^{p+1}\sigma\, \left[
e^{-\phi}\sqrt {|{\rm det} (g_{ij}+{\cal F}_{ij})|} + Ce^{{\cal F}} +
mI_{{\rm CS}}\right]\, ,
\end{equation}

\noindent where

\begin{equation}
g_{ij} = \partial_i X^\mu \partial_j X^\nu
G_{\mu\nu}\, ,
\end{equation}

\noindent is the metric on $\Sigma_{(p+1)}$ induced by the map $X$,
$(\mu, \nu=0, \dots 9)$ are the spacetime indices and ${\cal F}_{ij}\ 
(i=1,\cdots (p+1))$ is the modified 2-form field strength defined in
(\ref{calF}) ($B$ in ${\cal F}$ is the pull-back of the NS-NS 2-form
gauge potential $B$ with $X$). The second term in (\ref{action}) is a
WZ term where

\begin{equation}
C = \sum_{r=0}^{10} C^{(r)}
\end{equation}

\noindent is a formal sum of the RR gauge potentials $C^{(r)}$. 
It is understood that after expanding the potential only the
(p+1)-form is retained\footnote{Again here we have used the same
  symbols to denote the spacetime gauge potential and its pull-back
  with the map $X$.}.  The last term is only present for even p (the
IIA case) \cite{deroo}. Its coefficient $m$ is the cosmological
constant of massive IIA supergravity and $I_{{\rm CS}}$ is given in
\cite{ght}.

To construct supersymmetric D-p-brane actions, we replace the maps $X$
($\{X^\mu\}$) with supermaps $Z=(X,\theta)$ ($\{Z^M\}$) and the
various bosonic supergravity fields with the corresponding superfields
of which they are the leading component in a $\theta$-expansion.  The
frame index $A$ of the supervielbein decomposes under the action of
the D=10 Lorentz group as follows:

\begin{equation}
A = 
\left\{
\begin{array}{ccccc}
(a,\alpha) & a=0, \dots, 9\, , & & \alpha=1, \dots , 32\, , &
 \hspace{.5cm}{\rm for\,\, IIA} \\
& & & & \\
(a,I,\alpha) & a=0, \dots, 9\, , & I=1,2\, , & \alpha=1, \dots , 32\, , &
\hspace{.5cm}{\rm for\,\, IIB} \\
\end{array}
\right.
\end{equation}

\noindent where $a$ is a $d=10$ vector index and $\alpha$ is a $d=10$ 
spinor index in the Majorana representation. This notation allows to
treat the IIA and IIB theories in a unified way but it is understood
that in the IIB case chiral projection operators should be inserted in
appropriate places to reduce the Majorana spinor indices to
Majorana-Weyl ones.  The induced metric for both IIA and IIB D-branes
is

\begin{equation}
g_{ij} = E_i{}^aE_j{}^b\eta_{ab}\, ,
\end{equation}

\noindent where

\begin{equation}
E_i{}^A = \partial_i
Z^M E_M{}^A\, ,
\end{equation}

\noindent and $\eta_{ab}$ is the Minkowski (frame) metric. In what
follows, we shall assume that ${\rm det} \{g_{ij}\}\not=0$, unless
otherwise stated.

The action (\ref{action}) (including the fermions )
is invariant under the $\kappa$-transformations \cite{Be1}  

\begin{equation}
\label{kappa}
\left\{
\begin{array}{rcl}
\delta_{\kappa} Z^M E_M{}^a & = & 0\, ,\\
& & \\
\delta_{\kappa} Z^M E_M{}^\alpha  & = & [\bar \kappa(1+\Gamma)]^\alpha\, ,\\
& & \\
\delta_{\kappa} V_i &=& E_i{}^A\delta E^B B_{BA}\, ,\\
\end{array}
\right.
\end{equation}

\noindent with parameter $\kappa$.

The expression for $\Gamma$ for any D-p-brane is \cite{Be1}

\begin{equation}
\Gamma={\textstyle\frac{\sqrt {|g|}}{\sqrt{|g + {\cal F}|}}} \sum^\infty_{n=0} 
{1\over 2^n n!}
\gamma^{j_1k_1\dots j_nk_n}
{\cal F}_{j_1k_1}\dots {\cal F}_{j_nk_n} J^{(n)}_{(p)}\, , 
\label{gammaone}
\end{equation}

\noindent where $g = {\rm det}\left\{ g_{ij}\right\}$, 
$g+{\cal F}$ is shorthand for ${\rm det} (g_{ij}+ {\cal F}_{ij})$ and

\begin{equation}
J^{(n)}_{(p)}=\cases{\big(\Gamma_{11}\big)^{n+{p-2\over2}}
\Gamma_{(0)}\, ,\cr
\cr
(-1)^n(\sigma_3)^{n+{p-3\over 2}} i\sigma_2\otimes \Gamma_{(0)}\, .}
\label{gammatwo}
\end{equation}

\noindent The matrix $\Gamma_{(0)}$ is given by

\begin{equation}
\Gamma_{(0)}={\textstyle\frac{1}{(p+1)! \sqrt{|g|}}}
\epsilon^{i_1\dots i_{(p+1)}}
\gamma_{i_1\dots i_{(p+1)}} \, . 
\label{gammathree}
\end{equation}

Finally, the $32\times 32$ matrices $\gamma_i$ are defined as

\begin{equation}
\gamma_i = E_i{}^a \Gamma_a\, ,
\end{equation}

\noindent where $\{\Gamma_a; a=0,\dots 9\}$ are the spacetime 
gamma-matrices.  For later use, we note that

\begin{equation}
\left(\Gamma_{(0)} \right)^2 = (-1)^{{(p-1) (p-2)\over 2}}\, .
\label{gammafour}
\end{equation}

A crucial property of the $\kappa$-rules, which also plays an
important role in the actual proof of $\kappa$-invariance of the
D-p-brane actions, is that they must eliminate half of the fermionic
degrees of freedom. To see this, we remark that the total number of
bosonic physical degrees of freedom are $8$; $10-(p+1)$ are due to the
scalars $X$ and $p-1$ are due to the BI 1-form gauge potential $V$. In
order to make the number of physical bosonic degrees of freedom equal
to the number of fermionic ones, the $\kappa$-transformations must
eliminate exactly half of the fermionic degrees of freedom.  The
properties of the matrix $\Gamma$ defined in (\ref{inone}) ensure that
this is indeed the case: from $\Gamma^{2}=\II_{32\times 32}$ it
follows that all the eigenvalues of $\Gamma$ are $+1$ or $-1$. From
the tracelessness it follows that it has as many $+1$ as $-1$
eigenvalues, that is, it has $16$ of each, so the projector
$\frac{1}{2}(1+\Gamma)$ has 16 zero eigenvalues and 16 eigenvalues
equal to $+1$. This guarantees that $\kappa$-symmetry reduces the $32$
components of $\theta$ to $16$.  Due to the fact that the kinetic term
of $\theta$ in the BI action is linear in time derivatives there is a
second class constraint which reduces further the components of
$\theta$ from $16$ to $8$. Therefore the number of bosonic and
fermionic physical degrees of freedom are equal on the worldvolume.
The details of the invariance of the action (\ref{action}) under
$\kappa$-transformations and the proof that $\Gamma$ has the required
properties are given in, e.g.~\cite{Be1}.


\section{$\kappa$-Symmetry Revisited}

The main task in this section is to show that the product structure
$\Gamma$ associated with the D-p-brane $\kappa$-transformation law can
be written as given in (\ref{agamma}).
The proof is inspired by the work of \cite{BDL,BL} and is
similar for the IIA and IIB
D-p-branes. Because of this we shall present the IIA
case in detail and only the main points of the proof for the IIB case.
We begin by first rewriting the IIA product structure $\Gamma$ as

\begin{equation}
\label{susyq}
\Gamma={\textstyle\frac{\sqrt {|g|}}{\sqrt {|g+{\cal F}|}}}
{\rm se}^{{1\over2}{\cal F}_{jk} \gamma^{jk}
\Gamma_{11}}
 \Gamma^{\prime}_{(0)}\, ,
\end{equation}

\noindent where

\begin{equation}
{\rm se}^{{1\over2}{\cal F}_{jk}\gamma^{jk}}:=\sum^\infty_{n=0} {1\over 2^n n!}
\gamma^{j_1k_1\dots
j_nk_n} {\cal F}_{j_1k_1}\dots {\cal F}_{j_nk_n}\, ,
\end{equation}

\noindent so ``${\rm se}$'' stands for the skew-exponential function 
(i.e.~the usual exponential function with skew-symmetrized indices of
the gamma matrices at every order in the expansion so the expansion
has effectively only a finite number of terms), and

\begin{equation}
\Gamma^{\prime}_{(0)} = (\Gamma_{11})^{p-2\over2}\ \Gamma_{(0)}\, .
\end{equation}

\noindent It is worth noting that

\begin{equation}
\left(\Gamma_{(0)}^{\prime}\right)^2=1\, .
\end{equation}

To continue, we introduce a worldvolume (p+1)-bein, $e$, 
i.e.~$g_{ik}=e^{\ui}{}_i e^{\uk}{}_k \eta_{\ui\uk}$, where $\ui,
\uk=0,\dots, p$ are worldvolume frame indices. Then we rewrite
$\Gamma$ as

\begin{equation}
\Gamma={\textstyle\frac{1}{\sqrt {|\eta+{\cal F}|}}}
{\rm se}^{\frac{1}{2}X_{\ui\uk}
\gamma^{\ui\uk} \Gamma_{11}} \Gamma^{\prime}_{(0)}\, ,
\end{equation}

\noindent where ${\cal F}$ in the determinant is in the frame basis.
\noindent Then without loss of generality, we use
a worldvolume Lorentz rotation to write ${\cal F}$ as

\begin{equation}
{\cal F} \equiv {\textstyle\frac{1}{2}}
 {\cal F}_{\ui\uk} e^\ui\wedge e^\uk= \tanh\phi_0\, 
e^{0}\wedge e^{\ell}
+\sum_{r=1}^{\ell}\,\tan\phi_r\, e^{r}\wedge e^{ \ell+r}\, ,
\label{power}
\end{equation}

\noindent where $\{\phi_0, \phi_r; r=1,\dots, \ell\}$, 
$\ell= [p/2]$, are ``angles'' and $\{e^\ui\} =\{e^0, e^s;
s=1\dots, p\}$ is a Lorentz basis. Using this, we have

\begin{eqnarray}
\label{par2}
\sqrt {|\eta +{\cal F}|} & = & 
\biggl (-1+ \tanh^2 \phi_0 \biggr )^{1/2}
\Pi_{r=1}^\ell \biggl (1+\tan^2 \phi_r\ \biggr )^{1/2}
\nonumber\\
& & \nonumber \\
&=& {1\over \cosh\phi_0 \, \Pi_{r=1}^\ell\cos \phi_r}\, .
\end{eqnarray}

\noindent Substituting this in $\Gamma$, we get

\begin{equation}
\Gamma= \left(\cosh\phi_0 \, \Pi_{s=1}^\ell\, \cos
\phi_s\right)\,\,
{\rm se}^{\tanh\phi_0
\gamma^{\u0\ul}\Gamma_{11}}\,
\left(\Pi_{r=1}^\ell {\rm se}^{\tan\phi_r
\gamma^{\ur\ \url}\Gamma_{11}}\right)\,\, 
\Gamma_{(0)}^{\prime}\, .
\end{equation}

\noindent From the definition of ${\rm se}$, we can rewrite $\Gamma$ as

\begin{eqnarray}
\Gamma & = & 
\left(\cosh\phi_0+\sinh\phi_0\
\gamma^{\u0\ul}\Gamma_{11}\right) \times \nonumber\\
& & \nonumber\\
& & 
\left[\Pi_{r=1}^\ell \left(\cos\phi_r+\sin\phi_r\
\gamma^{\ur\ \url}\Gamma_{11}\right)\right]\,\, \Gamma_{(0)}^{\prime}\, ,
\end{eqnarray}

\noindent which in turn can be expressed as

\begin{eqnarray}
\label{rot}
\Gamma & = & 
e^{\phi_0 \gamma^{\u0\ul}\Gamma_{11}}\,\, 
\left(\Pi_{r=1}^\ell e^{\phi_r\gamma^{\ur\ \url} \Gamma_{11}}\right)\,\, 
\Gamma_{(0)}^{\prime}\\
& & \\
& = & 
\exp{\left\{\phi_0 \gamma^{\u0\ul}+\sum_{r=1}^\ell \phi_r
\gamma^{\ur\ \url}\Gamma_{11} \right\}}\Gamma_{(0)}^{\prime}\, .
\end{eqnarray}

In the last step we have used the fact that
$(\gamma^{\u0\ul}\Gamma_{11})^2=1$ while $(\gamma^{\ur\ \url}
\Gamma_{11})^2=-1$.  It is clear from this that the
product structure $\Gamma$ can be written as

\begin{equation}
\label{iiagamma}
\Gamma= e^{{1\over2} Y_{\ui\uk} \gamma^{\ui\uk}
\Gamma_{11}}\Gamma_{(0)}^{\prime}\, ,
\label{A}
\end{equation}

\noindent where

\begin{equation}
Y\equiv {\textstyle\frac{1}{2}} Y_{\ui\uk} e^\ui\wedge e^\uk=\phi_0\, e^0
\wedge e^\ell+\sum_{r=1}^\ell
\phi_r e^r\wedge e^{\ell+r}\, .
\end{equation}

Although we have shown this equation in a particular Lorentz frame, it
holds in any Lorentz frame. The relation between ${\cal F}$ and $Y$
is now

\begin{equation}
\label{relxy}
{\cal F}=``\tan'' Y\, ,
\end{equation}

\noindent where ``$\tan$'' is defined by eq.~(\ref{power}) in the special 
Lorentz frame. The explicit expression of the function ``$tan$'' is in
general frames more complicated but it can be always be found by going
to the special frame as an intermediate step.

Now let us turn to examine the product structure $\Gamma$ associated
with IIB D-p-branes. In this case the product structure $\Gamma$ can
be written as

\begin{equation}
\Gamma={\textstyle\frac{\sqrt {|g|}}{\sqrt {|g+{\cal F}|}}}
 {\rm se}^{-{1\over2} {\cal F}_{\ui\uk}
\sigma_3\otimes\gamma^{\ui\uk}}
\,\,\Gamma^{\prime}_{(0)}\, ,
\end{equation}

\noindent where

\begin{equation}
\Gamma^{\prime}_{(0)}=(\sigma_3)^{p-3\over2} i \sigma_2\otimes \Gamma_{(0)}\ ,
\end{equation}

\noindent is a ${\cal F}$-independent traceless product structure.

Following a similar computation as for the IIA D-p-branes, we find
that

\begin{equation}
\label{iibgamma}
\Gamma=e^{-{1\over2}Y_{\ui\uk} \sigma_3\otimes\gamma^{\ui\uk}}\
\Gamma^{\prime}_{(0)}\ ,
\end{equation}

\noindent where ${\cal F}$ and $Y$ are again related  as 
in eq.~ (\ref{relxy}).
 
Now, observing that $\Gamma_{(0)}^{\prime}$ anticommutes with the
gamma matrices that appear in the exponential in the expression for
$\Gamma$  we can write 

\begin{equation}
\Gamma= e^{-a/2}\ \Gamma_{(0)}^{\prime}\ e^{a/2}\, ,
\end{equation}

\noindent as in eq.~(\ref{agamma})of the introduction, where, as we
 have just shown

\begin{equation}
a=\left\{ 
\begin{array}{lc}
-{1\over2} Y_{jk} \gamma^{jk} \Gamma_{11}\, ,  \hspace{2cm} &{\rm IIA\, ,}\\
& \\
{1\over2} Y_{jk} \sigma_3\otimes\gamma^{jk}\, , &{\rm IIB\, .}
\end{array}
\right.
\end{equation}

As an application of the new expression for $\Gamma$ we remark that it
is straightforward to show that $\Gamma^2=1$ and ${\tr}\ \Gamma=0$
using the above-mentioned property of the exponential, the cyclic
properties of the trace and the analogous properties of
$\Gamma_{(0)}^{\prime}$.


\section{Supersymmetry}

We would like to derive here the condition (\ref{intwo}) of the
introduction for the fraction of supersymmetry preserved by a single
brane from the $\kappa$-symmetry transformation (\ref{inone}).  We
remark that the known $\kappa$-symmetry transformations of all M, IIA,
IIB, and heterotic branes have the same form, so the result of this
derivation applies to all these cases.  Here we consider the type II
case.  Since we are interested in {\sl bosonic} configurations that
preserve a fraction of the spacetime supersymmetry (i.e. we set
$\theta=0$ for these configurations), it is enough to examine the
supersymmetry transformation of the $\theta$ field up to terms linear
in $\theta$.  The supersymmetry and $\kappa$-symmetry transformations
of $\theta$ are

\begin{equation}
\label{transformation}
\delta \theta=(1+\Gamma)\kappa+\epsilon\ ,
\end{equation}

\noindent where $\epsilon$ is the space-time supersymmetry parameter.
Assuming a gauge fixing condition for $\kappa$-symmetry of the form

\begin{equation}
 {\cal P}\theta=0\ ,
\end{equation}

\noindent where ${\cal P}$ is a (field-independent) projection, 
${\cal P}^2 = {\cal P}$. The remaining non-vanishing components of
$\theta$ are given by $(1-{\cal P})\theta$ and the transformation
(\ref{transformation}) becomes a global supersymmetry transformation.
The condition for preserving the gauge-fixing condition ${\cal
  P}\delta\theta = (1-{\cal P})\delta\theta = 0$ is now equivalent to
having unbroken supersymmetry. Therefore, the condition for unbroken
supersymmetry is

\begin{equation}
\label{susya}
\delta\theta = \left(1+\Gamma\right)\kappa + \epsilon_{\rm unbr}=0 \, ,
\end{equation}

\noindent which in turn implies the condition 
$\left(1-\Gamma\right)\epsilon_{\rm unbr} = 0$ of the introduction.

For the IIA case, a convenient gauge-fixing condition is

\begin{equation}
\label{gfIIA}
\left(1+\Gamma_{11}\right)\theta = 0\, .
\end{equation} 

To obtain more explicit expressions we go to a (chiral) basis in which
$\Gamma_{11}$ is diagonal split the index $\alpha$ into the pair
$(\alpha_{1},\alpha_{2})$ with opposite chiralities,
$\alpha_{1},\alpha_{2}=1,\ldots,16$ so

\begin{equation}
\label{gamma11}
\Gamma_{11}= 
\left(
\begin{array}{cc}
\delta^{\alpha_{1}}{}_{\beta_{1}} & \\
& \\
 & -\delta^{\alpha_{2}}{}_{\beta_{2}} \\
\end{array}
\right)\, ,
\hspace{1cm}
\theta^{\alpha}=
\left(
\begin{array}{c}
\theta^{\alpha_{1}} \\
\\
\theta^{\alpha_{2}} \\
\end{array}
\right)\, .  
\end{equation}

\noindent and similarly for $\kappa^{\alpha}$ and $\epsilon^{\alpha}$.
In this basis the above gauge-fixing condition is simply
$\theta^{\alpha_{1}}=0$.  Since $\Gamma_{11}$ anticommutes with
$\Gamma$, in a basis that $\Gamma_{11}$ is diagonal, $\Gamma$ is
off-diagonal, i.e.

\begin{equation}
\Gamma =
\left( 
\begin{array}{cc}
& \Gamma^{\alpha_1}{}_{\beta_2}\\
& \\
\Gamma^{\alpha_2}{}_{\beta_1}& \\
\end{array}
\right)\, .
\end{equation} 

\noindent Preserving the  gauge-fixing condition 
$\theta^{\alpha_1} = 0$ in this basis $\delta\theta^{\alpha_1}= 0$
implies

\begin{equation}
\kappa^{\alpha_1} + \Gamma^{\alpha_1}{}_{\beta_2}\kappa^{\beta_2}
+ \epsilon^{\alpha_1} =  0\, .
\end{equation}

\noindent This leads to the (worldvolume) supersymmetry transformation

\begin{equation}
\delta\lambda^{\alpha_2} = -\Gamma^{\alpha_2}{}_{\beta_1}\epsilon^{\beta_1}
+\epsilon^{\alpha_2}\, ,
\end{equation}

\noindent where $\lambda = (1-{\cal P})\theta$.
For supersymmetric bosonic configurations $\delta\lambda^{\alpha_2} =
0$ as well, which is precisely the condition $(1-\Gamma)\epsilon = 0$.

For the IIB case it is convenient to choose as a gauge fixing condition

\begin{equation}
\label{gfIIB}
\left(1+\sigma_3\otimes \II_{32\times 32}\right)\ \theta = 0\, ,
\end{equation}

\noindent where $\theta$ is a doublet of chiral spacetime spinors. 
Again it is convenient to go to a basis in which $\sigma_3 \otimes
\II_{32\times 32}$ is diagonal, so

\begin{equation}
\sigma_3\otimes \II_{32\times 32} =
\left(
\begin{array}{cc}
\delta^{1,\alpha}{}_{1,\beta} & \\
& \\
& -\delta^{2,\gamma}{}_{2,\delta} \\
\end{array}
\right)\, ,
\hspace{1cm}
\theta = 
\left(
\begin{array}{c}
\theta^{1,\alpha} \\
\\
\theta^{2,\beta} \\
\end{array}
\right)\, ,
\end{equation}

\noindent and similarly for $\kappa^{A},\epsilon^{A}$. 
However, now we have to take into account the (positive) chirality of
the spinors. Thus, which the choice of $\Gamma_{11}$ matrix
(\ref{gamma11}) we split the spinors
$\theta^{1,\alpha},\theta^{2,\alpha}$ as in eq.~(\ref{gamma11})
and set to zero the negative chirality components 
$\theta^{1,\alpha_{2}},\theta^{2,\alpha_{2}}$ so

\begin{equation}
\theta^{1,\alpha}=
\left(
\begin{array}{c}
\theta^{1,\alpha_{1}} \\
\\
0 \\
\end{array}
\right)\, ,  
\hspace{1cm}
\theta^{2,\alpha}=
\left(
\begin{array}{c}
\theta^{2,\alpha_{1}} \\
\\
0 \\
\end{array}
\right)\, .  
\end{equation}

\noindent (The same applies to the spinors $\kappa,\epsilon$.) 

In this basis, the gauge-fixing condition is simply
$\theta^{1,\alpha_{1}}=0$. Again, $\sigma_3\otimes \II$ anticommutes
with $\Gamma$. Therefore, in the above basis that $\sigma_3 \otimes
\II$ is diagonal, $\Gamma$ is off-diagonal as in the IIA
case\footnote{Here we have restricted already $\Gamma$ to the
  positive-chirality subspace.}:

\begin{equation}
\Gamma=
\left(
\begin{array}{cc}
 & \Gamma^{\delta_{1}}{}_{\beta_{1}} \\
& \\
\Gamma^{\gamma_{1}}{}_{\alpha_{1}} \\
\end{array}
\right)\, .  
\end{equation}

\noindent The supersymmetry transformation is given by

\begin{equation}
\delta\lambda^{2,\alpha_{1}} = \epsilon^{2,\alpha_{1}}
-\Gamma^{\alpha_{1}}{}_{\beta_{1}}\epsilon^{1,\beta_{1}}\, ,
\end{equation}

\noindent where $\lambda = (1-{\cal P})\theta$.

It is instructive to compare this supersymmetry transformation with
the one of the supersymmetric $d=10$ Maxwell theory in Minkowski
space.  For this, we have to linearize the 9-brane supersymmetry
transformation in terms of the BI field. This leads to

\begin{equation}
\delta\lambda^{2,\alpha_{1}} = \epsilon^{2,\alpha_{1}} - 
\epsilon^{1,\alpha_{1}} - F_{ij}[\gamma^{ij}]^{\alpha_{1}}{}_{\beta_{1}}
\epsilon^{1,\beta_{1}}\, ,
\end{equation}

\noindent which reproduces the supersymmetry transformation
of the usual Maxwell theory with parameter $\epsilon^{2,\beta}$ when
$\epsilon^{1,\alpha}=\epsilon^{2,\alpha}$ as well as
Volkov-Akulov-type supersymmetries.

Finally, we note that the conditions (\ref{gfIIA}) and (\ref{gfIIB})
are {\it covariant} gauge-fixing conditions for the $\kappa$-symmetry.
This is rather different from the type IIA/IIB fundamental string
which is plagued with a well-known covariant quantization problem. The
reason why this distinction between the type IIA/IIB fundamental
string and the D-p-branes occurs is explained in more detail in
Appendix A.


\section{Supersymmetric D-Brane Probes}

Let us consider a single D-brane probe propagating in $d=10$ Minkowski
spacetime. The field equations of the probe are

\begin{eqnarray}
\partial_i \left\{ \sqrt {|g+F| }\left[ (g+F)^{-1}\right]^{(ij)}
\partial_j X^M\right\} &=&0\, ,\nonumber\\
& & \\
\partial_i \left\{ \sqrt {|g+F|}\left[(g+F)^{-1}\right]^{[ij]}
 \right\} &=&0\, .\nonumber 
\end{eqnarray}

A solution of these equations is 

\begin{equation}
\left\{
\begin{array}{rcl}
X^i &=& \sigma^i\, , i=0,1,\cdots ,p\, ,\\
& & \\                     
X^m &=& y^m\, ,m=p+1,\cdots , 9\, ,\\
& & \\
F_{ij} &=& c_{ij}\, ,\\
\end{array}
\right.
\end{equation}

\noindent where $y^m$ are the positions of the probe and $c_{ij}$ 
are constant.

As we have seen in the previous section the condition for the above
configuration to be supersymmetric is

\begin{equation}
\label{susysusy}
 \left(1-e^{-a/2}\ \Gamma^{\prime}_{(0)}\ e^{a/2}\right)\epsilon=0\, , 
\end{equation}

\noindent where

\begin{equation}
a=\cases{-{1\over2} Y_{jk} \gamma^{jk} \Gamma_{11}\qquad\qquad &{\rm
IIA\, ,}\cr
&\cr
{1\over2} Y_{jk} \sigma_3\otimes\gamma^{jk}\qquad\qquad &{\rm IIB\, .}}
\end{equation}

\noindent Viewing the D-p-brane as
a (p+1)-dimensional Minkowski subspace of $d=10$ Minkowski spacetime,
it is clear due to the properties of $\Gamma$ this configuration
preserves $1/2$ of the supersymmetry of the $d=10$ Minkowski vacuum.

Next suppose that two D-brane probes with non-vanishing but constant
BI field are placed in the $d=10$ Minkowski spacetime with product
structures $\Gamma$ and $\tilde \Gamma$.  It is rather involved
to find the fraction of the supersymmetry preserved by a generic such
configuration. Below we shall examine some special
cases. 


\subsection{Orthogonal Intersections}

Suppose that two D-branes, with product structure $\Gamma$ and
$\tilde\Gamma$, respectively, are intersecting orthogonally, and that
both BI field strengths are zero. Because of the latter hypothesis,
$a=0$, and so

\begin{equation}
\left\{
\begin{array}{rcl}
\Gamma & = &\Gamma^{\prime}_{(0)}\, , \\
& & \\
\tilde\Gamma&=&\tilde\Gamma^{\prime}_{(0)}\, .\\
\end{array}
\right.
\end{equation}

Viewing the two D-branes as (p+1)-and (q+1)-dimensional Minkowski
subspaces of the $d=10$ Minkowski spacetime, one can introduce an
orthonormal basis $\{e_a; a=0,\dots,9\}$ in the $d=10$ target space
adopted to these two D-branes, i.e. the orthonormal basis is chosen by
extending an orthonormal basis along the common directions of the
intersection first along the relative transverse directions of the
intersection and then along the overall transverse directions of the
intersection, in the terminology of \cite{gppkt}.  Using the $d=10$
gamma-matrices adopted to this orthonormal basis,
$\Gamma^{\prime}_{(0)}$ and $\tilde\Gamma^{\prime}_{(0)}$ can be
expressed as a product of gamma-matrices and therefore

\begin{equation}
\Gamma\tilde\Gamma=\pm\tilde\Gamma\Gamma\ ,
\end{equation}

If $\Gamma$ and $\tilde\Gamma$ commute, they can be diagonalized
simultaneously and their product $\Gamma\tilde\Gamma$ is also a
product structure.  If $\Gamma \ne \tilde \Gamma$, then
$\Gamma\tilde\Gamma$ is traceless, ${\rm tr}\Gamma\tilde\Gamma=0$, in
which case the amount of supersymmetry preserved is $1/4$.  Examples
of orthogonally intersecting D-brane configurations preserving $1/4$
of the supersymmetry are those with 4 or 8 relative transverse
directions in agreement with \cite {poly,qgreen}. If $\Gamma = \tilde
\Gamma$, then the two D-p-branes are parallel and the fraction of
supersymmetry preserved is $1/2$.

However, if $\Gamma$ and $\tilde\Gamma$ anticommute, imposing
(\ref{susysusy}) separately for each D-brane leads to the breaking of
{\it all} spacetime supersymmetry (however, see also Ref.~\cite{gptown}).


\subsection{Branes Intersecting at Angles}

Another special case is that of two intersecting D-p-branes at an
arbitrary angle in $d=10$ Minkowski spacetime with the 2-form BI field
vanishing \cite{BDL,BL,BaL}\footnote{For recent results on 
supergravity solutions that are related to 
branes at angles, see \cite{ggpt,a1,a2,a3,a4}.}. 
For each D-brane involved in the configuration, we can
associate a $d=10$ Lorentz frame; we may assume without loss of
generality that the two orthonormal frames coincide along the
directions of the intersection.  For this, we use the assumption that
each brane is identified with a Minkowski subspace of the $d=10$
Minkowski spacetime to choose an orthonormal basis for the worldvolume
directions and then extend this basis to an orthonormal basis for the
whole $d=10$ Minkowski spacetime. If $\{e_a; a=1\dots 10\}$ is the
Lorentz frame associated with the first D-brane and $\{\tilde e_a;
a=1\dots 10\}$ is the Lorentz frame associated with the second
D-brane, there is a Lorentz transformation $\Lambda$ such that

\begin{equation}
\label{basis}
\tilde e_a=  e_b\,\Lambda^b{}_a\ .
\end{equation}

This in turn implies that the gamma-matrices $\{\Gamma_a;a=1,\dots,
10\}$ in the frame $\{e_a; a=1\dots 10\}$ are related to the
gamma-matrices $\{\tilde\Gamma_a; a=1,\dots, 10\}$ in the frame
$\{\tilde e_a; a=1\dots 10\}$ as follows:

\begin{equation}
\label {spinor}
\tilde \Gamma_a= \Gamma_b\, \Lambda^b{}_a\ =S^{-1}\, \Gamma_a\, S\ .
\end{equation}

\noindent where $S$ is an element in $Spin(1,9)$ that 
depends on $\Lambda$.  As in the previous case of parallel or
orthogonal D-branes,

\begin{eqnarray}
\Gamma&=&\Gamma^{\prime}_{(0)}\, ,\\
&&\nonumber\\
\tilde\Gamma&=&\tilde\Gamma^{\prime}_{(0)}\, .
\end{eqnarray}

\noindent However, in this case $\Gamma$ is a product of the 
gamma-matrices associated with the $\{e\}$ basis while $\tilde\Gamma$
is a product of the gamma-matrices associated with the $\{\tilde e\}$
basis.  Using (\ref{spinor}), the latter product structure written in
the $\{e\}$ basis is

\begin{equation}
\label{basis2}
\tilde \Gamma_{(\tilde e)}= S^{-1}\tilde\Gamma_{(e)} S\ ,
\end{equation}

\noindent where the subscript denotes the basis with respect to 
which $\tilde \Gamma$ is expressed and $\tilde\Gamma_{(e)}$ is again a
product of gamma-matrices in the $\{e\}$ basis.  Dropping the
subscript and expressing both supersymmetry projection operators in
the $\{e\}$ basis, we get

\begin{eqnarray}
\label{ssusy}
\Gamma\epsilon&=&\epsilon\, ,\\
&&\nonumber\\
\label{ssusy2}
S^{-1}\tilde\Gamma S \epsilon&=&\epsilon\, .
\end{eqnarray}

The case $\Gamma=\tilde \Gamma$ and $S\not=1$ was studied in
\cite{BDL,BL,BaL}} where it was shown that the fraction of the
supersymmetry preserved is ${k/32}$, where $k$ is the number of
singlets of the matrix $S$ acting on the spinors $\epsilon$ that have
the property, $\Gamma\epsilon=\epsilon$.  We remark that such
intersecting at an angle configuration of two D-p-branes is not
associated with Lorentz rotations $\Lambda$ of the worldvolume
coordinates of a single D-p-brane. This is because from the definition
of the product structures $\Gamma=\tilde\Gamma$ and $S=1$, so 
condition (\ref{ssusy2}) is not independent from condition (\ref{ssusy})
and the supersymmetry preserved is $1/2$.  Therefore the interesting
cases involve Lorentz rotations of the $d=10$ spacetime that are not
Lorentz rotations of the worldvolume coordinates of a single
D-p-brane. In fact the relevant Lorentz rotations are those of the
relative transverse coordinates of the intersecting configuration, in
the terminology of \cite{gppkt}.  Examples of Lorentz rotations that
have singlets acting on $SO(1,9)$ spinors are those that lie in the
subgroups $SU(n)$, $1\leq n\leq 3$, $Sp(2)$, $G_2$ and $Spin(7)$ of
$SO(1,9)$.

Next suppose that $\Gamma\not=\tilde \Gamma$ and $S\not=1$, then since
both $\Gamma$ and $\tilde\Gamma$
are products of $d=10$ gamma-matrices $\Gamma$ and $\tilde \Gamma$
either commute or anti-commute.  If they commute, there is a basis
that can be simultaneously diagonalized, i.e.

\begin{eqnarray}
\Gamma&=& (\delta^{\alpha_1}{}_{\beta_1},\delta^{\alpha_2}{}_{\beta_2},
-\delta^{a_1}{}_{b_1}, -\delta^{a_2}{}_{b_2})\, ,\cr
&&\nonumber\\
\tilde\Gamma&=&(\delta^{\alpha_1}{}_{\beta_1},-\delta^{\alpha_2}{}_{\beta_2},
\delta^{a_1}{}_{b_1}, -\delta^{a_2}{}_{b_2})\, .
\end{eqnarray}
In this basis, only the spinors $\epsilon=(\epsilon^{\alpha_1},
\epsilon^{\alpha_2}, 0,0)$ satisfy 
(\ref{ssusy}). Substituting this $\epsilon$ into (\ref{ssusy2}), we get

\begin{eqnarray}
S^{\beta_2}{}_{\alpha_1}
\epsilon^{\alpha_1}+S^{\beta_2}{}_{\alpha_2}\epsilon^{\alpha_2}&=&0\, ,
\\
&&\nonumber\\
S^{a_2}{}_{\alpha_1}
\epsilon^{\alpha_1}+S^{a_2}{}_{\alpha_2}\epsilon^{\alpha_2}&=&0\ .
\end{eqnarray}

If ${\rm det}(\{ S^{\beta_2}{}_{\alpha_2}\})\not=0$, the first
equation can be solved for $\epsilon ^{\alpha_2}$ and after substitution
into the second equation we get

\begin{equation}
A^{a_2}{}_{\alpha_1}\epsilon^{\alpha_1}=0\, ,
\end{equation}

\noindent where

\begin{equation}
\label{matrixA}
A^{a_2}{}_{\alpha_1}=S^{a_2}{}_{\alpha_1}-S^{a_2}{}_{\alpha_2}
(S^{-1})^{\alpha_2}{}_{\beta_2}S^{\beta_2}{}_{\alpha_1}\ .
\end{equation}

\noindent Thus the fraction of the supersymmetry preserved is $k/32$ 
where $k$ is the number of zero eigenvalues of the matrix $A$. Note
that, since  ${\rm tr}\Gamma \tilde\Gamma=0$, $A$ is an $8\times 8$ square
matrix.  The case ${\rm det}(\{ S^{\beta_2}{}_{\alpha_2}\})=0$ can be
treated in a similar way.

Now if $\Gamma$ and $\tilde \Gamma$ anticommute, there is a basis such
that

\begin{eqnarray}
\Gamma & = & 
\pmatrix{{\II} & 0 \cr
        0& -{\II}}\, , \nonumber \\
& & \\
\tilde \Gamma(e)&=&\pmatrix{0 & U \cr
        U & 0 }\, , \nonumber 
\end{eqnarray}

\noindent where ${\II}$ is a $16\times 16$ unit square matrix 
and $U$ is a diagonal $16\times 16$ matrix with $U^2={\II}$. Note
that there is a matrix $V$ such that

\begin{equation}
\tilde \Gamma=V^{-1} D V\, ,
\end{equation}

\noindent where $D$ is a diagonal matrix with $D^2={\II}$ and 
${\rm tr} D=0$.  Since now $D$ and $\Gamma$ commute, we can examine
this case by repeating the steps of the previous case after setting
$S\rightarrow VS$. In particular, the fraction of the supersymmetry
preserved is $k/32$ where $k$ are the number of zero eigenvalues of
the matrix $A$ defined in (\ref{matrixA}), after replacing $S$ with $VS$.


\subsection{Branes Intersecting at Angles with BI Fields}

It remains to investigate the case of intersecting D-branes with
non-vanishing constant BI field $F$. As we have seen in the
previous section the effect that a non-vanishing BI field has on the
supersymmetry projection of a D-brane is to rotate it. In this respect
the situation is similar to the one examined above but there is an
important difference. {\sl The rotation induced by the BI field on the
  D-brane product structure is a relative rotation of the 
left and right moving fields}. More explicitly, we deduce
from the form of the
exponential in (\ref{iiagamma}) (IIA) and
(\ref{iibgamma}) (IIB) that the dependence of the BI field is
induced by a Lorentz rotation that acts differently on the
two 16-component (left and right moving) $\kappa$-symmetry
parameters. In the IIA case this is due to the fact that
the exponential has
a $\Gamma_{11}$ that multiplies the standard generator of Lorentz
rotations in the spinor representation.  In the IIB case the different
behaviour of the left and right moving fields is due to the presence of
the $\sigma_3$ matrix in the exponential.

Intuitively, it is clear that the dependence on the BI field
cannot be written as a Lorentz rotation that acts the
same on the left and right moving fields. We recall that the BI field is 
a non-linear generalization of
the Maxwell field on the worldvolume of the D-branes. Now if the only
effect that it has is to induce a Lorentz rotation, it would mean that
by changing Lorentz frame one could set the BI field equal to zero.
This would have been a contradiction since the Maxwell equations are
Lorentz invariant and if the Maxwell field is non-zero in one Lorentz
frame it is non-zero in any Lorentz frame. 

Nevertheless, the supersymmetry conditions for two intersecting
D-branes in $d=10$ Minkowski spacetime can be written as

\begin{eqnarray}
\label{general}
\Gamma\ e^{a/2}\ \epsilon&=&e^{a/2} \epsilon\, ,
\cr
&&\cr
e^{-\tilde a/2}\ \tilde \Gamma\ e^{\tilde a/2}\epsilon & = & \epsilon\, .
\end{eqnarray}

\noindent Now if the two D-branes intersect at a angle, then as before we
introduce two $d=10$ Lorentz frames one for each D-brane.  Then there is
a $d=10$ Lorentz transformation, $\Lambda$, as in (\ref{basis2}) that
relates the $d=10$ gamma-matrices adopted to one frame to the
gamma-matrices adopted to the other frame.  Rewriting (\ref{general})
in the same basis, we get

\begin{eqnarray}
\label{gena}
\Gamma^{\prime}_{(0)} e^{a/2}\ \epsilon & = & e^{a/2}\ \epsilon\, ,
\cr
&&\cr
S^{-1}\  e^{-\tilde a/2}\ \tilde \Gamma^{\prime}_{(0)}\
 e^{\tilde a/2}\ S\ \epsilon&=&
\epsilon\, ,
\end{eqnarray}

\noindent where $S$ is induced by $\Lambda$ and $\Gamma^{\prime}_{(0)}$ 
and $\tilde \Gamma^{\prime}_{(0)}$ are expressed in the same basis. Next, let
us set

\begin{equation}
\eta = e^{a/2}\ \epsilon\ .
\end{equation}

\noindent Then (\ref{gena}) can be rewritten as

\begin{eqnarray}
\label{genb}
\Gamma^{\prime}_{(0)}\ \eta & = & \eta\, ,
\\
&&\nonumber\\
\label{genb2}
T^{-1}\ \tilde \Gamma^{\prime}_{(0)}\ T\ \eta & = & \eta\ ,
\end{eqnarray}

\noindent where

\begin{equation}
T=e^{ \tilde a/2}\ S\ e^{-a/2}\ .
\end{equation}

\noindent To investigate the fraction of supersymmetry preserved 
by two D-branes intersecting at angles with non-vanishing BI fields,
we remark that (\ref{genb}), (\ref{genb2}) is the same as
(\ref{ssusy}), (\ref{ssusy2}) after setting

\begin{equation}
T\rightarrow S\ .
\end{equation}

Therefore the methods developed in the previous subsection to
examine the fraction of supersymmetry preserved by two intersecting
D-branes at an angle without BI fields also apply to this case.


\section{Supersymmetric M-brane probes}

The $\kappa$-symmetry transformation for the M-5-brane, in the
form given in \cite{Ho2}, is

\begin{equation}
\delta\theta= (1+\Gamma)\kappa\ ,
\end{equation}

\noindent where

\begin{eqnarray}
\Gamma&=&\Gamma_{(0)}  \left[1-{1\over 2\cdot 3!} h_{ijk} \gamma^{ijk}
\right]\, ,
\label{M5g-1}\\
&&\nonumber\\
\Gamma_{(0)}&=& {\textstyle\frac{1}{6! \sqrt{|g|}}} \epsilon^{i_1\cdots i_6}
\gamma_{i_1}\cdots
\gamma_{i_6}\ .\label{M5g0}
\end{eqnarray}

\noindent The metric $g$ is the induced metric and $h$ is a self-dual 3-form
worldvolume field.  Observe that $\Gamma$ and
$\Gamma_{(0)}$ are traceless hermitian product structures.

As we have already mentioned in section 2, the supersymmetry preserved
by a M-5-brane probe is

\begin{equation}
(1-\Gamma)\epsilon=0\, ,
\end{equation}

\noindent where $\epsilon$ is the supersymmetry parameter.
As in the case of D-branes, $\Gamma$ can be written as

\begin{equation}
\label{rotmbrane}
\Gamma=e^{-a}\Gamma_{(0)}= e^{-{1\over2}a}\Gamma_{(0)} e^{{1\over2}a}\, ,
\end{equation}

\noindent where

\begin{equation}
a=-{\textstyle\frac{1}{2\cdot 3!}} h_{ijk} \gamma^{ijk}\ .
\end{equation}

Note that $a^2=0$ due to the selfduality of $h$. 

Although the product structure $\Gamma$ is easily written in the form
given above, the dependence on $h$ has no straightforward geometric
interpretation as a kind of rotation since the exponential in
(\ref{rotmbrane}) is cubic in the worldvolume gamma-matrices.
Nevertheless, one can repeat the analysis of section 3 to find the
fraction of supersymmetry preserved by a configuration of intersecting
M-5-brane probes. A single M-5-brane probe preserves $1/2$ of the
supersymmetry of D=11 vacuum with or without a non-vanishing $h$.

To investigate the supersymmetry preserved by two intersecting
M-5-brane probes at an angle with constant field $h$, we can again
proceed as in the case of D-branes.  For this, we introduce two
Lorentz frames, $\{e\}$ and $\{\tilde e\}$ adopted to each M-5-brane
involved in the intersection, and a D=11 Lorentz transformation,
$\Lambda$, such that $\tilde e_a\,=\, e_b\, \Lambda^b{}_a$.  Then,
there is a $S\in Spin(1,10)$, $S=S(\Lambda)$, such that

\begin{equation}
\tilde{\Gamma}_{(0)}= S^{-1}\ \Gamma_{(0)}\ S\ .
\end{equation}

\noindent Using this, we can write both supersymmetry conditions
in the same D=11 gamma-matrix basis as follows:

\begin{eqnarray}
\Gamma_{(0)}\ e^{a/2}\ \epsilon & = & e^{a/2}\ \epsilon\, ,\\
&&\nonumber\\
S^{-1}\ e^{-\tilde a/2}\ \Gamma_{(0)}\ e^{\tilde a/2}\ S\ \epsilon &=&
\epsilon\, .
\end{eqnarray}

\noindent Next setting,

\begin{equation}
\eta=e^{a/2}\epsilon\ ,
\end{equation}

\noindent the supersymmetry conditions can be rewritten as

\begin{eqnarray}
\Gamma_{(0)}\ \eta & = & \eta\, ,\\
&&\nonumber\\
T^{-1}\  \Gamma_{(0)}\ T\ \eta &=& \eta\ ,
\end{eqnarray}

\noindent where

\begin{equation}
T=e^{\tilde a/2} S e^{-a/2}\, ,
\end{equation}

\noindent which is of the form (\ref{genb}), (\ref{genb2}). 
Because of this, the general analysis for D-branes applies in this
case as well and so we shall not repeat it here.

As an example let us consider the intersection of two M-5-branes on a
string \cite{ggpt}. Let us suppose that $a=\tilde a=0$. The rotation
$\Lambda$ involved in this case is an element of $Sp(2)\subset
SO(1,10)$. The spinor representation of $SO(1,10)$ decomposed as
representation of $Sp(2)$ has $6$ singlets and the fraction of the
supersymmetry preserved by such configuration is $3/16$.

This method of finding the fraction of supersymmetry preserved by
intersecting M-5-branes configurations can be easily extended to
intersecting configurations involving M-2-branes as well.  For
example, the supersymmetry conditions for a M-2-brane/M-5-brane
intersecting configuration at an angle are

\begin{eqnarray}
\Gamma_{(0)}\ \epsilon &=&\epsilon\, ,\cr
&&\nonumber\\
S^{-1}\ e^{-\tilde a/2}\  \tilde \Gamma_{(0)}\ e^{\tilde a/2}\  S\
\epsilon &=& \epsilon\, ,
\end{eqnarray}

\noindent where

\begin{equation}
\Gamma_{(0)}={\textstyle\frac{1}{3!\sqrt {|g|}}}
\epsilon^{i_1i_2i_3}\gamma_{i_1}\gamma_{i_2}\gamma_{i_3}
\end{equation}

\noindent is the product structure associated with the M-2-brane,
and $\tilde \Gamma_{(0)}$ is the product structure associated with the
M-5-brane as given in (\ref{M5g-1}), (\ref{M5g0}) (both expressed in
the same basis).  Finally, the supersymmetry conditions for a
M-2-brane/M-2-brane intersecting configuration at an angle are

\begin{equation}
\left\{
\begin{array}{rcl}
\Gamma_{(0)}\ \epsilon & = &\epsilon\, ,\\
& & \\
S^{-1}\ \Gamma_{(0)}\ S\ \epsilon  & = & \epsilon\, ,
\end{array}
\right.
\end{equation}

\noindent where $\Gamma_{(0)}$ is the product structure associated
with one of the M-2-branes.


\section{Supergravity Backgrounds}

We briefly consider the coupled D-brane/supergravity equations.  The
supergravity solution corresponding to D-p-branes in the string frame
is

\begin{equation}
\left\{  
\label{sugra}
\begin{array}{rcl}
ds^2&=& H^{-{1\over2}} ds^2(\bE^{(1,p)})+H^{1\over2}ds^2(\bE^{9-p})\, ,
\\
& & \\
e^\phi&=& H^{3-p\over4}\, ,\\
& & \\
F_{p+2}&=&\omega(\bE^{(1,p)})\wedge d H^{-1}\, ,\\
\end{array}
\right.
\end{equation}

\noindent where $\omega$ is the volume form of $\bE^{(1,p)}$ and 

\begin{equation}
H=H(y-y_0)
\end{equation}

\noindent is a harmonic function of $\bE^{(9-p)}$.
The embedding $X$ of the worldvolume into the spacetime is specified
by identifying the worldvolume coordinates of the D-p-brane with
$\bE^{(1,p)}$, i.e.

\begin{equation}
\label{embedding}
\left\{
\begin{array}{rcl}
X^i & = & \sigma^i\, , \\
& & \\
X^m & = & y^m_0\, ,  \\
\end{array}
\right.
\end{equation}

\noindent where $y_0$ is the position of the harmonic function $H$.
It is straightforward to see that the BI field must vanish by
examining the field equation of the NS/NS 2-form gauge potential.

It remains to compare the supergravity Killing equation with the
worldvolume supersymmetry condition (\ref{intwo}).  The solution of
the Killing spinor equation is

\begin{equation}
\epsilon = H^{-{1\over8}} \xi\, ,
\end{equation}

\noindent for constant  $\xi$ and

\begin{equation}
(1-\Pi)\ \xi = 0\, ,
\label{regularframe}
\end{equation}

\noindent where

\begin{equation}
\Pi=
\left\{
\begin{array}{lr}
(\Gamma_{11})^{p-2\over2}\Gamma_0\cdots \Gamma_{p}\, , 
& {\rm IIA}\, ,\\
 & \\
(\sigma_3)^{p-3\over2} i\sigma_2\otimes \Gamma_0\cdots \Gamma_{p}\, , 
& {\rm IIB}\, .\\
\end{array}
\right.
\end{equation}

\noindent Using the solution  (\ref{sugra}) and (\ref{embedding}), 
we find that $\Gamma=\Pi$.

It is natural to extend the above analysis to the case of intersecting
brane configurations. For orthogonally intersecting ones, we find that
the BI field must vanish for each brane separately. This is also the
case for all intersecting brane configurations (even for those that
intersect at angles) with vanishing NS/NS 2-form gauge potential.

Finally, as for the single brane the conditions for supersymmetry
derived from the supergravity Killing spinor equations are compatible
with those found in section 5.


\section*{Acknowledgments}

We thank M.~Douglas for helpful discussions. G.P.~thanks the Physics
Department of Stanford University and the institute for Theoretical
Physics of Groningen University for hospitality during the course of
this work, and the Royal Society for financial support.  E.B.~and
T.O.~would like to express their gratitude to the Physics Department
of Stanford University for the stimulating working environment and
hospitality.  The work of E.B.~ is supported by the European
Commission TMR programme ERBFMRX-CT96-0045, in which he is associated
to the University of Utrecht.  The work of R.K.~was supported by the
NSF grant PHY-9219345.  The work of E.B.~R.K.~and T.O.~has also been
supported by a NATO Collaboration Research Grant.  T.O.~would also
like to thank M.M.~Fern\'andez for her support.


\appendix


\section{$\kappa$-Symmetry and Covariant Quantization}

The $\kappa$-symmetry is an example of an {\it infinitely reducible}
gauge symmetry because the sequence of shifts $\kappa_0 \rightarrow
(1-\Gamma)\kappa_1$, $\kappa_1\rightarrow (1+\Gamma)\kappa_2\, \cdots$
leaves the $\kappa$-transformations (\ref{kappa}) unchanged.  This is
a property of all branes. However, as far as the covariant
quantization is concerned, there is a distinction between the type II
fundamental string and the D-branes.  This is because the only
covariant gauges that we can pick are

\begin{equation}
\left\{
\begin{array}{rcll}
\Gamma_{11}\theta & = & 
\pm \theta\, ,\hskip 1truecm & {\rm IIA}\, ,\\
& & \\
\sigma_3\otimes \II_{32\times 32} \ \theta  & = & 
\pm \theta\, ,& {\rm IIB}\, .
\end{array}
\right.
\end{equation}

Under $\kappa$-symmetry the variation of the action is given by the
following expression \cite{Ce1,Ag1,Be1}

\begin{equation}
\delta_{\kappa} S =  \int d^{p+1} \sigma \;  
\delta_{\kappa} \bar \theta  \Delta
 = \int d^{p+1} \sigma \;  \bar \kappa (1+\Gamma)  \Delta =0\, .
\end{equation}
Here  $\Delta = (1-\Gamma) \Psi$. This variation can be presented
as consisting
of two parts: one due to the variation of $ \bar \theta {\cal P}$ and the
other, due to variation of $ \bar \theta (1- {\cal P})$:
\begin{equation}
\delta_{\kappa} S =  \delta_{\kappa}^1 S + \delta_{\kappa}^2 S=
\int d^{p+1}
\sigma \;  \bar \kappa (1+\Gamma)  {\cal P} \Delta +  \int d^{p+1}
\sigma \;
\bar \kappa (1+\Gamma) (1- {\cal P} )\Delta\, .
\end{equation}

Assume that we choose the gauge $\bar \theta {\cal P} =0$ and do not
vary $\bar \theta {\cal P}$ anymore.  The variation of the action
under the transformations of the remaining part of $\theta$, which is
given by $\delta \bar \theta (1- {\cal P}) $ should not vanish
anymore. This would mean that the gauge symmetry is gauge-fixed, the
action does not have a symmetry anymore.  Thus we have to find whether

\begin{equation}
\delta_{\kappa} S_{\rm g.f.}  =    \delta_{\kappa}^2 S =     \int d^{p+1}
\sigma \;  \bar \kappa (1+\Gamma) (1- {\cal P} )\Delta
\end{equation}

\noindent vanishes or not. We observe that

\begin{equation}
\delta_{\kappa} S_{\rm g.f.}  =     \int d^{p+1} \sigma \;  \bar \kappa
(1+\Gamma) (1- {\cal P} )(1-\Gamma) \Psi\, .
\end{equation}

This explains why the issue of commutativity (non-commutativity) of
the projector ${\cal P}$ with $\Gamma$ becomes so important. Indeed if
they commute

\begin{equation}
[ {\cal P},\, \Gamma]=0\, ,
\end{equation}

\noindent the action  still has a local symmetry since

\begin{eqnarray}
\delta_{\kappa} S_{\rm g.f.}   & = &     \int d^{p+1} \sigma \;  \bar
\kappa
(1+\Gamma) (1- {\cal P} )(1-\Gamma) \Psi \nonumber \\
& & \nonumber \\
& = & \int d^{p+1} \sigma \;  \bar \kappa
(1+\Gamma)(1-\Gamma)  (1- {\cal P} )\Psi =0\, ,
\end{eqnarray}

\noindent and therefore $\bar \theta {\cal P} =0$ is not an admissible gauge
condition.  However, if they anticommute

\begin{equation}
\{ {\cal P},\, \Gamma\}=0\, ,
\end{equation}

\noindent then

\begin{eqnarray}
\delta_{\kappa} S_{\rm g.f.}   & = &     \int d^{p+1} \sigma \;  \bar
\kappa (1+\Gamma) (1- {\cal P} )(1-\Gamma) \Psi \nonumber \\
& & \nonumber \\
& = & - \int d^{p+1} \sigma \;  \bar
\kappa (1+\Gamma) (1+\Gamma)  {\cal P}  \Psi \neq 0\, ,
\end{eqnarray}

\noindent the action is not gauge-symmetric anymore, the gauge-fixing 
condition is admissible.

For the fundamental GS string the $\kappa$-symmetry is given by

\begin{equation}
\delta  \bar \theta = \bar \kappa (1+ \Gamma)\, ,  \qquad   \Gamma =
\Gamma_{(0)}
\quad   {\rm at }  \quad p=1\, .
\end{equation}

\noindent This expression for $\Gamma$ is proportional to  
$\Gamma^a \Gamma^b$ and therefore it commutes with ${\cal P} = {1\over
  2} (1+ \Gamma_{11})$ in the IIA case and with ${\cal P} = {1\over 2}
(1+ \sigma_3\otimes \II )$ in the IIB case. These would be Lorentz
covariant gauges for the fundamental string, and as we see here, they
are not acceptable. This is the well known covariant quantization
problem of the IIA/IIB fundamental string.

On the other hand we have found that for all D-branes the relevant
$\Gamma$ anticommute with Lorentz covariant gauge-fixing projectors
above.  The gauges are acceptable since the remaining action is not
gauge symmetric.  This explains the existence of the covariant gauges
for D-branes which were found in \cite{Ag1}.


\end{document}